\documentclass{svjour3} 
\smartqed

\usepackage{graphicx}

\usepackage{mathrsfs}
\usepackage{bm}
\usepackage{bbm}
\usepackage{xr}
\usepackage{amsmath}
\usepackage{amssymb}
\usepackage{amsfonts}
\usepackage{graphicx}
\usepackage{lscape}

\usepackage[dvips]{color}

\providecommand*{\I}{\mathrm{i}}                           %% imaginary unit i
\providecommand*{\ket}[1]{|#1\rangle}                      %% ket vector
\providecommand*{\rbra}[1]{(#1|}                           %% bra with parenthesis
\providecommand*{\rket}[1]{|#1)}                           %% ket with parenthesis
\providecommand*{\lrbra}[1]{\rbra{\widetilde{#1}}}    %% left bra (with tilde) with parenthesis
\providecommand*{\mrmd}{d}									 %% differential ``d'' in mathrm
\newcommand*{\umat}[1]{\underline{\mathscr{#1}}}      %% underline script symbol, used for matrices
\renewcommand{\vec}[1]{\bm{#1}}                       %% vectors are bold symbols here
\renewcommand{\thesection}{\arabic{section}}
\renewcommand{\thesubsection}{\arabic{section}.\arabic{subsection}}
\renewcommand{\thetable}{\arabic{table}}

\begin{document}

\title{\Large Longitudinal Atomic Beam Spin Echo Experiments: A possible way to study Parity Violation in Hydrogen}

\author{Maarten~DeKieviet
\and
Thomas~Gasenzer
\and
Otto~Nachtmann
\and
Martin-I.~Trappe
}

\institute{M. DeKieviet \at
              Physikalisches Institut, Universit{\"a}t Heidelberg, Philosophenweg 12, 69120 Heidelberg,\\
              Tel.: +123-45-678910\\
              Fax: +123-45-678910\\
              \email{maarten@physi.Uni-Heidelberg.DE}       
           \and
           T. Gasenzer \at
              Institut f{\"u}r Theoretische Physik, Universit{\"a}t Heidelberg, Philosophenweg 16, 69120 Heidelberg, Germany
           \and
           O. Nachtmann \at
              Institut f{\"u}r Theoretische Physik, Universit{\"a}t Heidelberg, Philosophenweg 16, 69120 Heidelberg, Germany
           \and
           M.-I. Trappe \at
              Institut f{\"u}r Theoretische Physik, Universit{\"a}t Heidelberg, Philosophenweg 16, 69120 Heidelberg, Germany
}

\date{Received: date / Accepted: date}

\maketitle

\abstract{
We discuss the propagation of hydrogen atoms in static electric and magnetic fields in a longitudinal atomic beam spin echo (lABSE) apparatus.
Depending on the choice of the external fields the atoms may acquire both dynamical and geometrical quantum mechanical phases.
As an example of the former, we show first in-beam spin rotation measurements on atomic hydrogen, which are in excellent agreement with theory.
Additional calculations of the behaviour of the metastable 2S states of hydrogen reveal that the geometrical phases may exhibit the signature of parity-(P-)violation. This invites for possible future lABSE experiments, focusing on P-violating geometrical phases in the lightest of all atoms.\\[2ex]
\hfill
{\small HD--THEP--10--05}
 % end of PACS codes
} %end of abstract

%========================================================================================
\section{Longitudinal Atomic Beam Spin Echo using Ground State Atomic Hydrogen} \label{s:exp}
%========================================================================================

\begin{figure}[htb]
\centering
\includegraphics[scale=0.485]{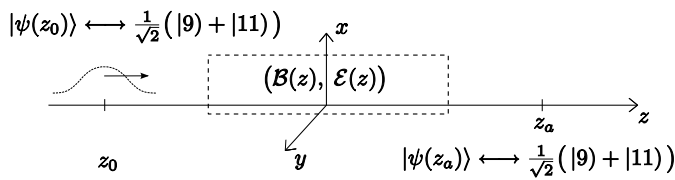}
\caption{Schematic view of the longitudinal spin echo interferometer including the definition of the coordinate system. The coordinate along the beam direction is denoted by $z$. The atom is prepared in a wave packet around $z_0$ and analysed around $z_a$. In the example discussed in Section \ref{s:theo} we start with a superposition $\ket{\psi(z_0)}$ of two states, labeled with $\rket{9}$ and $\rket{11}$. After passing the magnetic and electric fields $\vec{\mathcal B}(z)$ and $\vec{\mathcal E}(z)$ the wavefunction is projected onto an analysing state $\ket{\psi(z_a)}$, for example again onto a superposition of the states $\rket{9}$ and $\rket{11}$.}
\label{SchematiclABSE}
\end{figure}

In this contribution we discuss a possible experimental route to measuring P-violating effects, based on the methods available with an atomic beam spin echo interferometer as described in \cite{ABSE95}. 
While many theoretical and experimental studies of P-violation due to neutral current exchange focus on heavy atoms (e.g. \cite{Khrip91,Bou97,BeWi99,Vet95} and recently \cite{tsigutkin-2009}), the experimental observation of P-violation in the lightest of atoms is still an open problem \cite{DuHo07}. In \cite{BeGaNa07_I,BeGaNa07_II} P-violating geometrical phases were studied which may provide a novel route to measuring the weak neutral current effects in hydrogen.
Since atom interfero\-meters are phase sensitve, we choose a longitudinal atomic beam spin echo (lABSE) setup of the type described in \cite{ABSE95}, with which dynamical quantum mechanical phases have already been successfully measured for a variety of light atomic species in their electronic ground state  \cite{ABSE95,ZiStSchDeGr98,DeHaRe06}.  The analogous scheme of an apparatus suitable for metastable hydrogen atoms is  shown in Figure \ref{SchematiclABSE}. Herein, we consider the hydrogen atom being in a superposition of $n=2$ internal states and travelling through an interferometer consisting of static electric and magnetic fields acting on the atom.
The magnetic field strength $\vec{\mathcal B}$ and the electric field strength $\vec{\mathcal E}$ are supposed to depend only on the $z$ coordinate and to be nonzero only between $z_0$ and $z_a$. Here $z_0$ marks the place at the beginning of the interferometer around which, in a field-free region, the wave packet of the atom is prepared. Similarly, $z_a$ at the end of the interferometer marks the place where, again in a field-free region, the internal state of the atom is analysed. For hydrogen state preparation and analysis is done through the insertion of a multipole Stern-Gerlach filter which  transmits only those atoms being in the desired hyperfine state. 

\begin{figure}[htb]
\centering
\includegraphics[scale=0.3]{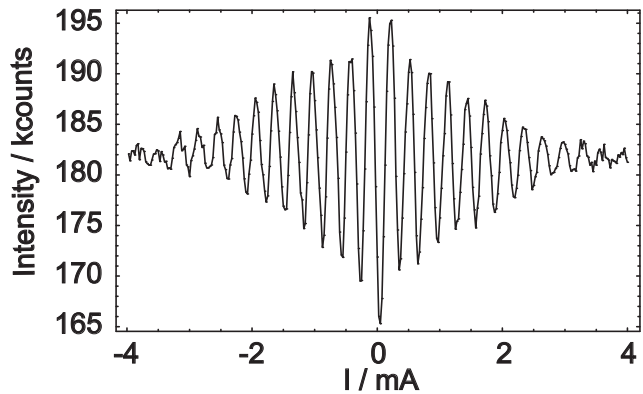}
\caption{Experimental spin rotation curve for an atomic hydrogen beam along $z$ in its $n=1$ electronic ground state. The electric field 
$\vec{\mathcal E} =0$. The current $I$ through a single cylindrical coil and, correspondingly, the magnetic field $\mathcal B_z$ along $z$ are varied (i.e. ${\mathcal B}_x={\mathcal B}_y=0$). The polariser and analyser are configured antiparallel with respect to $x$.}
\label{H-lABSE-Data}
\end{figure}

So far, a typical atomic beam spin echo experiment consists of measuring the count rate, i.e. the intensity of the overall transmitted beam, while $\vec{\mathcal E}(z)=0$, and two subsequent, antiparallel magnetic coils are driven with variable currents. 
In its simplest form only a single coil is in use and the spin echo curve degenerates to a so-called spin rotation curve. In Fig. \ref{H-lABSE-Data} we show a spin-rotation curve, obtained with a beam of H atoms in their $n=1$ ground state. These data are acquired with the polariser and analyser being antiparallel, which corresponds to the situation in which the $\rket{1S_{1/2},1,\,1}_x$ state is prepared and the $\rket{1S_{1/2},1,-1}_x$ state is analysed. The subscript $x$ denotes the quantisation axis. Therefore the spin rotation curve is centered around a global minimum for the applied current $I = 0$. 

Using the same apparatus, not only spin rotation measurements, but also first in-beam spin echo experiments were performed on ground state atomic hydrogen and deuterium. The corresponding results will be published elsewhere  \cite{DeHaRe06}. For the first species, one can observe three spin echo groups,
while there are five for a deuterium beam. The quantum mechanical treatment of the experiment describes the location of the spin echo groups in dependence of
the spin echo current and is in excellent agreement with the data. 
 
These experiments demonstrate two major advantages of lABSE over conventional NMR spectroscopy. Due to the in-beam arrangement the probing particles are in a state of very high rarefaction (i.e. are essentially non-interacting within the beam), and their internal temperature is essentially $T\approx 0$. This makes it possible to achieve in this atom interferometer an energy resolution in the peV range! In addition, the detected signal is not due to the nuclear magnetic moment (leading to an extremely small energy quantum) but rather due to the magnetic moment of the entire particle. This becomes an appreciable advantage when applying the lABSE technique to metastable atoms, like the $n=2$ state in hydrogen discussed below.

%========================================================================================
\section{Longitudinal Atomic Beam Spin Echo using Metastable Hydrogen}\label{s:theo}
%========================================================================================

In the following we will consider the evolution of the meta\-stable 2S states of atomic hydrogen within this lABSE setup and calculate the relevant phases which atoms pick up when travelling through the interferometer. A concrete proposal for observables particularly sensitive to P-violating geometrical phases in an lABSE experiment will be presented elsewhere.
If not indicated otherwise, we use natural units $\hbar=c=1$ and the notation explained in \cite{BeGaMaNaTr08_I}.

The Schr\"odinger equation describing the undecayed $n=2$ states of hydrogen reads
\begin{align}\label{2.2}
\I\frac{\partial}{\partial t}|\psi(\mathbf x, t)\rangle=
\left[-\frac{1}{2m}\Delta +{\umat{M}}(z)+E_0\right]|\psi(\mathbf x,t)\rangle\ .
\end{align}

The -- in general non-Hermitian -- mass matrix $\umat{M}(z)$ incorporates the P-violating 2S-2P mixing effects due to the Z-boson exchange between the electron and the proton, and the couplings of the atom to the external electric and magnetic fields. For each $z$, $\umat{M}(z)$ has, in general, 16 linearly independent right and, analogously, left eigenvectors with complex energies 
\begin{align}\label{2.9}
\umat{M}(z)\rket{\alpha(z)}=E_\alpha(z)\rket{\alpha(z)}\ ,\;\; \lrbra{\alpha(z)}\umat{M}(z)=\lrbra{\alpha(z)}E_\alpha(z)\ ,\;\;(\alpha=1,\dots,16)\ .
\end{align}

In the following we consider (\ref{2.2}) in the adiabatic limit. That is, we choose conditions where the evolution of each metastable 2S state ($\alpha=9,\dots,12$ in our notation) decouples from that of all the other $n=2$ states. We assume that the atoms enter the apparatus with longitudinal momentum $k\gg \sqrt{2m|E_\alpha(z)|}$. The solution of (\ref{2.2}) for the 2S states can then be constructed using an extension of the WKB method  \cite{BeGaNa07_I,BeGaNa07_II,BeGaMaNaTr08_I}. The wave functions contain, as expected from \cite{Ber84}, dynamical and geometrical phase factors with the geometrical phases defined as
\begin{align}\label{2.32}
\gamma_{\alpha\alpha}(z)=\I\int^z_{z_0}\mrmd z'\,
\lrbra{\alpha(z')}\frac{\partial}{\partial z'}\rket{\alpha(z')}\ ,\; (\alpha=9,\dots,12)\ .
\end{align}
P-violating effects enter to first order in the geometrical but only to second order in the dynamical phases. Thus, a measurement of the geometrical phases gives a new way to study atomic P-violation.

Typically, phases are measured in interference experiments. As an example we consider a Gaussian wave packet of a coherent superposition of two 2S states entering the apparatus, $\ket{\psi(z_0)}=\frac{1}{\sqrt{2}}(\rket{9}+\rket{11})$, where $\rket{9}$ and $\rket{11}$ denote the states that evolve adiabatically from the empty-space 2S states with $F=1$, $F_3=+1$ and $-1$, respectively. We let these states run through the apparatus where they are split due to the magnetic field and come together again at the analysis point $z=z_a$. There we project onto the initial superposition of the states $\rket 9$ and $\rket{11}$ and suppose that the total time-integrated flux ${\cal F}_p$ of atoms in that projection is measured. We have calculated ${\cal F}_p$ for a wave packet of mean longitudinal momentum $\bar k_m$ and width $\sigma'_k$ and find
\begin{align}\label{4.31}
{\cal F}_p=\sum_{\alpha,\beta\in \{9,11\}}\frac14 \exp [-(\Delta\tau_\beta-\Delta\tau_\alpha)^2/(8\sigma'^2_k)]\, U^*_\beta(z_a,z_0;\bar k_m)\,U_\alpha(z_a,z_0;\bar k_m)\ .
\end{align}
Here $U_\alpha(z_a,z_0;\bar k_m)=\exp[-\I\phi_{\mathrm{dyn},\alpha}+\I\phi_{\mathrm{geom},\alpha}]$ contains the dynamical and geometrical phases. Up to small corrections we get, with $m$ the atom's mass, $\phi_{\mathrm{dyn},\alpha}=(m/\bar k_m)\int^{z_a}_{z_0}\mathrm dz'\,E_\alpha(z')$ and $\phi_{\mathrm{geom},\alpha}=\gamma_{\alpha\alpha}(z_a)-\gamma_{\alpha\alpha}(z_0)$. The reduced arrival times of the two parts ($\alpha=9$ and $11$) of the wave packet at $z=z_a$ are
\begin{align}\label{4.25}
\Delta\tau_\alpha=\frac{m}{\bar k^2_m}\int^{z_a}_{z_0}\mathrm dz'\,\mathrm{Re}E_\alpha(z')\ .
\end{align}
In the experiment the flux ${\cal F}_p$ is measured as function of the magnetic field in the analysing part of the apparatus. This field is scaled by a factor $s$ varying typically between $0.8$ and $1.2$.

\begin{figure}[htb]
\centering
\includegraphics[scale=0.9]{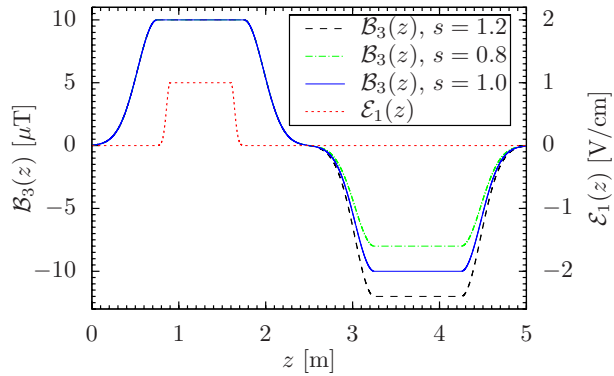}
\caption{The magnetic field component $\mathcal B_3(z)$ and the electric field component $\mathcal E_1(z)$. The parameter $s$ is a measure of the detuning of the second half of the magnetic field configuration around the antisymmetric arrangement.}
\label{5MeterFelderNEU}
\end{figure}

\begin{figure}[htb]
\centering
\includegraphics[scale=0.9]{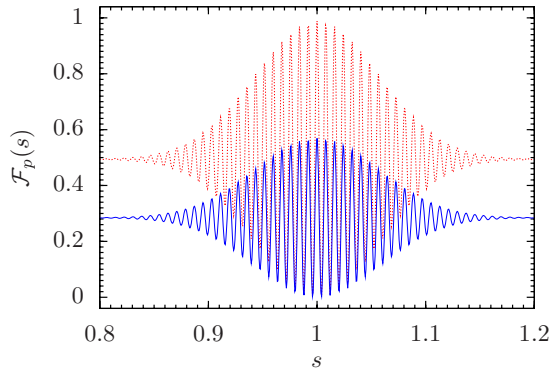}
\caption{The spin echo signal corresponding to the case where only the magnetic field is present is given by the (red) dashed line. The spin echo signal corresponding to the presence of the magnetic field and the electric field is given by the solid line.}
\label{Spinecho_all_script}
\end{figure}

Let us now consider a configuration with magnetic and electric fields as shown in Figure \ref{5MeterFelderNEU} to discuss our result (\ref{4.31}). In Figure \ref{Spinecho_all_script} we show the calculated signal, ${\cal F}_p(s)$, for the cases $\mathcal E=0$ and $\mathcal E\not= 0$. In both cases the geometrical phases $\phi_{\mathrm{geom},\alpha}$ are zero. For $\mathcal E=0$ we see a clear oscillation pattern due to the different variation with $s$ of the dynamical phases $\phi_{\mathrm{dyn},9}$ and $\phi_{\mathrm{dyn},11}$. The envelope of the interference signal is governed by the width $\sigma'_k$ of the Gaussian wave packet.

For $\vec{\mathcal E}\not= 0$ the Stark effect leads to a shift of the complex energies $E_\alpha$. The shift of the real parts of $E_\alpha$ turns out to be negligible. Therefore, the reduced arrival times do not change substantially compared to the case of $\vec{\mathcal E}= 0$. The change of the oscillation frequency of $\mathcal F^{\vec{\mathcal E}\not=\vec 0}_p(s)$ is not significant either if we compare to $\mathcal F^{\vec{\mathcal E}=\vec 0}_p(s)$, see Figure \ref{Spinecho_all_script}. However, the electric field admixes 2P to the 2S states, thereby making them decay faster. That is, the decay widths $\Gamma_\alpha=-2\,\mathrm{Im}E_\alpha$ with $\alpha=9,11$ are increased leading to a smaller total integrated flux via the dynamical phases.

In future work we will study field configurations where geometrical phases, both P-conserving and P-violating ones, occur. In principle the spin echo experiments should provide the experimental access to these geometrical phases since a non-vanishing geo\-metrical phase will modify signals like those of Figure \ref{Spinecho_all_script}. For example, switching on suitable electric fields, geometrical phases will contribute to the total phase factors $U_\beta^*$ and $U_\alpha$ in (\ref{4.31}). Besides the changed magnitudes of the peaks, the crucial information is the displacement of their positions in the $s-\mathcal F_p(s)$ interference diagram compared to a reference signal indicating non-zero geometrical phases. Of course, a P-violating geometrical phase will modify the reference signal in a slightly different way if the measurement is performed with the space-reflected field configuration. In this way, the high sensitivity of lABSE experiments could possibly lead to measurements of P-conserving as well as P-violating geometrical phases in metastable hydrogen.

\bibliographystyle{spphys}
\bibliography{myapvbib_20091001}

\end{document}